%% file: Vehlhaber.Salazar.ECC26.tex
\documentclass[a4, 10 pt, conference]{ieeeconf} 
\pdfoutput=1 
\IEEEoverridecommandlockouts
\usepackage[english]{babel}
\usepackage{cite}
\usepackage{amsmath,amssymb,amsfonts}
\usepackage{algorithm}
\usepackage{algorithmic}
\usepackage{graphicx}
\usepackage{textcomp}
\usepackage{dsfont}
\usepackage[usenames,dvipsnames]{xcolor}
\usepackage{units}
\usepackage{booktabs}
\usepackage{comment}
\usepackage{url}
\usepackage[normalem]{ulem}
\usepackage[hidelinks]{hyperref}
\usepackage{caption}
\usepackage{subcaption}
\usepackage{flushend}
\usepackage{lettrine}
\usepackage{lipsum}

\usepackage{tikz}
\usetikzlibrary{arrows.meta,
	positioning,shapes,shadows,calc,external}
\usepackage{placeins}

\newcounter{probcounter}
\newtheorem{prob}[probcounter]{Problem}

\newtheorem{theorem}{Theorem}[section]

\renewenvironment{prob}[1]{\noindent\textbf{Problem \theprobcounter} (#1) \it \stepcounter{probcounter}}{}

\newcommand{\vsp}{-10pt} 
\newcommand{\tM}{t_\mathrm{M}}
\newcommand{\posM}{p_\mathrm{M}}

\def\BibTeX{{\rm B\kern-.05em{\sc i\kern-.025em b}\kern-.08em
    T\kern-.1667em\lower.7ex\hbox{E}\kern-.125emX}}

\newif\ifmargincomments 
\margincommentsfalse

\ifmargincomments

\else

\fi

\newif\ifrev
\revtrue 

\input{def}

\renewcommand{\baselinestretch}{0.99}

\begin{document}
\title{\LARGE \bf
	A Model Predictive Control Scheme for Flight Scheduling\\ and Energy Management of Electric Aviation Networks
}
\author{Finn Vehlhaber and Mauro Salazar
\thanks{Control Systems Technology section, Department of Mechanical Engineering, Eindhoven University of Technology, Eindhoven, The Netherlands
        {\tt\small \{f.n.vehlhaber,m.r.u.salazar\}@tue.nl}}%
}
\maketitle
\thispagestyle{empty}
\pagestyle{empty}

\begin{abstract}
This paper presents a Model Predictive Control (MPC) scheme for flight scheduling and energy management of electric aviation networks, where electric aircraft transport passengers between electrified airports equipped with sustainable energy sources and battery storage, with the goal of minimizing grid dependency.
	Specifically, we first model the aircraft flight and charge scheduling problem jointly with the airport energy management problem, explicitly accounting for local weather forecasts.
	Second, we frame the minimum-grid-energy operational problem as a mixed-integer linear program and solve it in a receding horizon fashion, where the route assignment and charging decisions of each aircraft can be dynamically reassigned to mitigate disruptions.
	We showcase the proposed MPC scheme on real-world data taken from a conventional flight network and weather conditions in the US American North East.
	The proposed framework saves between 10 and 37~\% of grid energy requirements when compared to a baseline without re-routing. Hence, results show that MPC can effectively guarantee operation of the network by efficiently re-assigning flights and rescheduling aircraft charging, while maximizing the efficiency of the on-site energy systems. 
\end{abstract}


\input{sections/intro.tex}
\input{sections/methodology.tex}
\input{sections/results.tex}

\input{sections/conclusion.tex}
\section{Acknowledgments}
We thank  Dr.\ I.\ New for proofreading this paper and D.~Fernández\ Zapico for helping with the NSRDB API.
\vspace{-5pt}
\bibliographystyle{IEEEtran}
\bibliography{../../../bibliography/main.bib,../../../bibliography/SML_papers.bib,../../../bibliography/aviation.bib}

\end{document}

%% file: def.tex
\ifmargincomments

\else

\fi

\ifrev

\else

\fi

\newcommand{\flexbrac}[1]{\if\relax\detokenize{#1}\relax \else (#1) \fi}
\newcommand{\flexcomma}[1]{\if\relax\detokenize{#1}\relax \else ,#1 \fi}



%% file: sections/intro.tex
\section{Introduction}
For decades, airports have been providing essential mobility to remote communities and served as their connection to the rest of the world. Flights scheduled in these networks are often performed by small general aviation aircraft, that could be replaced with first generation electric aircraft in the near future. With the growing diffusion of these aircraft into regional aviation markets, however, electric power requirements may soon exceed the local supply and have an adverse impact on the communities these flights serve. One idea to mitigate this trend is to take advantage of the large premises available at airports and install renewable energy generation and storage systems in order to promote grid independence.
Equipped with such local clean power grids, regional airports may then find an additional purpose in their local ecosystem, and can serve as regional energy hubs~\cite{AntcliffBorerEtAl2021}. Their battery energy storage system (BESS) would be a valuable energy buffer for the local power grid, and ultimately enable airports to function as virtual power plants to provide demand response. 

Control schemes are essential for said operation, especially given the intermittent nature of renewable energy sources and the uncertainty in power demand.
What is more, airports must still be able to serve their main purpose---that is, to ensure the flight operations. Through electric flights, airports form a coupled energy-transportation network, which has to be carefully managed. Specifically, the inability to charge an aircraft at a certain destination airport at a future time may be attenuated by instead charging more at the current airport if an excess of power is available. Such strategic dynamic charge scheduling and fleet reassignment that adapts to current network disruptions can therefore significantly increase the energy and operational efficiency and reduce the burden on power systems.

Against this backdrop, we devise a model predictive control (MPC) scheme to dynamically assign electric aircraft to regional commuter flights, schedule their charging, and control the state of charge in the BESS installed at each airport in the network.
\begin{figure}[t!]
	\centering
	\input{sections/energyModelFigure.tikz}
	\caption{Energy model of an airport $h\in\mathcal{H}$ with renewable energy sources in addition to the grid connection and a stationary battery (BESS). Arrows indicate positive direction of power flow.\label{fig:airport}}
\end{figure}
\paragraph*{Related Literature}
This paper is related to two streams of research, namely aircraft routing and real-time charge scheduling. The aircraft assignment problem has been widely studied in the past, with objectives ranging from fleet minimization, to efficiently meeting maintenance requirements or reducing fuel expenditure~\cite{Barnhart1998,Guerkan2016}, and was often framed as a (mixed-)integer linear program (MILP/ILP). These assignments are usually computed whenever an airline implements a new seasonal flight schedule, and can reach very large dimensions, so they are often solved sequentially~\cite{Barnhart2004}. For smaller time horizons, authors have successfully leveraged network-flow formulations on time-extended digraphs~\cite{Hane1995,Roy2007}. Fügenschuh et al. used such an approach for the scheduling and routing of safari planes, where they also included the necessity for refueling~\cite{Fuegenschuh2013}. Zeghal et al. implement a tailored solution approach for a flexible aircraft routing problem that meets the large demand variability for a specific airline~\cite{Zeghal2011}. Naturally, most of the literature solves these problems for conventional aircraft, but recent publications have also addressed routing of electrified fleets. Here, some authors focus on the network design~\cite{Justin2022,Kinene2023}, while others solve the problem in conjunction with the infrastructure optimization~\cite{Mitici2022,Trainelli2021}. The growing reliability on renewable energy sources has also been considered in this context~\cite{VehlhaberSalazar2024,Amstel2023}, and it has been shown that accounting for these in the schedule can result in regional airport's grid independence, albeit potentially degrading the level of service~\cite{VehlhaberSalazar2023b}. While the aircraft routing problem for electric fleets has been addressed in the literature, also with the consideration of intermittent energy sources, we believe that a real-time assignment approach has not yet been investigated. To this end, MPC offers compelling benefits to address the fleet assignment problem in a receding-horizon framework.

MPC has found success in various scheduling problems in transportation networks due to the framework's ability to address schedule disruptions and changes in other conditions in real-time~\cite{TsaoMilojevicEtAl2019,CavoneBoomEtAl2022}. Implemented on a real electric bus network, one such dynamic fleet management approach was shown to effectively mitigate system disturbances in real-time, thus reducing operational costs~\cite{RinaldiPicarelliEtAl2019}. It was also widely used in vehicle-to-grid applications~\cite{LeFlochDiMeglioEtAl2015,ZhengSongEtAl2019}, where it was capable of successfully providing demand response while guaranteeing sufficient vehicle charge. Applied to flight assignment for on-demand urban air mobility, it was shown to significantly reduce delay when compared to a na\"ive assignment approach~\cite{KleinbekmanMiticiEtAl2019}. To the best of our knowledge, however, the problem has not yet been studied for vehicle scheduling in electric multi-hub transportation networks, and especially not in the context of electric flying.

\paragraph*{Statement of Contributions}
In this paper we study the benefits of real-time assignment and scheduling of an electric aircraft fleet to serve regional flight demands while reducing the dependence on local power grids through the optimal use of locally harvested renewable energy. For this purpose, we devise an MPC scheme that dynamically re-schedules aircraft and flights to adapt to locally varying conditions im a regional aviation network.
\paragraph*{Notation} In the following we often concatenate variables into column vectors denoted with bold letters as in $\mathbf{v} = [v(1)\; v(2)\; \dots \;v(n)]^\top$, such that the dimension of $\mathbf{v}$ would be $n\times1$. A column vector of all ones with dimension $m$ is denoted $\mathbf{1}_m$, the $m\times m$ identity matrix is denoted $I_m$, and $\Lambda_m$ denotes an $m\times m$ lower triangular matrix of ones. Matrices are denoted with capital letters $M$, and the element in the $m$-th row and $n$-th column is indexed with $[M]_{mn}$. We use $\otimes$ for the Kronecker product operator.

%% file: sections/energyModelFigure.tikz
\begin{tikzpicture}[
	node distance = 6mm and 12mm,
	B/.style = {draw, minimum width=12mm,fill=white},
	arr/.style = {draw=black, line width= 1pt, -{Triangle[fill=black]}},
	N/.style =  {circle,minimum width=5pt,thick,draw,inner sep=2pt,fill=white},
	Thor/.style = {anchor=south},
	Tver/.style = {anchor=west},
	]
	\node (bat) [B,general shadow={fill=black,shadow xshift=.5mm,shadow yshift=-.5mm},minimum height=10mm] {\parbox[b][10mm]{12mm}{\centering \large BESS}};
	\draw[thick] ($(bat.east)+(0,-0.075)$) to [out=180,in=270] ($(bat.north) + (-0.075,0)$) ;
	\node [below left=0mm and 0mm of bat.north east]{$E_\mathrm{b}^h$};
	\node (n) [N,right=of bat.east] {};
	\node (apron) [B,right=of n.east,minimum height=20mm,minimum width=30mm] {\parbox[b][20mm]{12mm}{\large apron}};
	\node (charger) [draw,fill=black,minimum height=14mm,minimum width=1mm,inner sep=0pt, above right= -4mm and 12mm of n.east] {};
	\node (grid) [above=8mm of n.north,draw=black,circle,minimum width=1mm] {};
	\node (solar) [draw,minimum width=7mm,minimum height = 4mm,shading = axis, left color=BlueViolet, right color=RoyalBlue,shading angle=135,general shadow={fill=Blue,shadow xshift=.5mm,shadow yshift=1mm},below=of n.south] {};
	\node[draw=black,circle,minimum width=1mm] at (grid.north) {}; 
	\begin{scope}[shift={($(charger.east)+(14mm,-8mm)$)},scale=0.5]
		\draw[thick,fill=white] (0,2) -- (-1.7,1.8) -- (-1.7,1.6) -- (0,1.5) -- (1.7,1.6) -- (1.7,1.8) -- cycle;
		\draw[thick,fill=white,x radius = .6,y radius=.1] (0,.2) circle;
		\draw[thick,rounded corners=5,fill=white] 
		(.05,.05) .. controls (0,0) .. (-.05,.05)
		(-.05,.05) .. controls  (-0.7,2) and (-.1,3) .. (0,3.2)
		.. controls (.1,3) and (.7,2) .. (.05,.05);
		\node (p1) at (0,1.9) [inner sep=0pt] {\tiny$E_\mathrm{b}^1$};
		\draw ($(p1.north west)+(.05,.2)$) to [out=340,in=90] ($(p1.east)+(0,.1)$);
		\draw ($(p1.east)+(0,-.15)$) to [out=270,in=0] ($(p1.south west)+(0,-.15)$);
	\end{scope}
	\begin{scope}[shift={($(charger.east)+(25mm,-12.5mm)$)},scale=0.5]
		\draw[thick,fill=white] (0,2) -- (-1.7,1.8) -- (-1.7,1.6) -- (0,1.5) -- (1.7,1.6) -- (1.7,1.8) -- cycle;
		\draw[thick,fill=white,x radius = .6,y radius=.1] (0,.2) circle;
		\draw[thick,rounded corners=5,fill=white] 
		(.05,.05) .. controls (0,0) .. (-.05,.05)
		(-.05,.05) .. controls  (-0.7,2) and (-.1,3) .. (0,3.2)
		.. controls (.1,3) and (.7,2) .. (.05,.05);
		\node (p2) at (0,1.9) [inner sep=0pt] {\tiny$E_\mathrm{b}^k$};
		\draw ($(p2.north west)+(.05,.2)$) to [out=340,in=90] ($(p2.east)+(0,.1)$);
		\draw ($(p2.east)+(0,-.15)$) to [out=270,in=0] ($(p2.south west)+(0,-.15)$);
	\end{scope}
	\draw[arr] (bat) edge node[Thor] {$P_\mathrm{b}$} (n);
	\draw[arr] (n) to node[Thor] {$P_\mathrm{a}$} (n -| charger);
	\draw[arr] (grid) edge node[anchor=south west] {$P_\mathrm{gr}$} (n);
	\draw[arr] (solar) edge node[Tver] {$P_\mathrm{rnw}$} (n);
	\draw[arr] (charger.east |- p1.north west) to node[Thor] (p1label) {$P_\mathrm{c}^1$} (p1.north west);
	\draw[arr] (charger.east |- p2.north west) to node[anchor=north east] {$P_\mathrm{c}^{k}$} (p2.north west);
\end{tikzpicture}

%% file: sections/methodology.tex
\section{Methodology}
In this section we devise an MPC scheme for the real-time control and scheduling of the aircraft fleet. First, we introduce the model, inspired by our previous work~\cite{VehlhaberSalazar2023b}, for the considered time horizon. Then, we define the objective function with the MPC-typical terms and state the receding horizon optimization problem. Finally, we describe an approach to simulate the flights to generate measurement inputs for the controller.

\subsection{Real-time Control Problem}
We adopt an MPC control scheme, where at each time sample we collect measurements of the state variables of our system and compute optimal input trajectories for an upcoming horizon based on these inputs. Our system consists of a network of regional airports that make up the set $\mathcal{H}$ between which electric flights are operated, the schedule of which is known beforehand and the scheduled flights for the day are collected in the set $\mathcal{F}$, where each flight is a tuple $f=(t_f^\mathrm{sd},\hat{t}_f,o_f,d_f,\hat{E}_f)$ of the scheduled departure time~$t_f^\mathrm{sd}$, the estimated flight time~$\hat{t}_f$, the origin~$o_f$ and destination~$d_f$, and the estimated energy expenditure~$\hat{E}_f$ for the flight. Said flights are assigned to aircraft in the fleet $\mathcal{P}$ by our controller, whereby the assignment is decided in real-time to adapt to the current deviations from the estimated conditions of the system, namely changes in the flight time and energy expended during flights, as well as deviations from the weather forecast that influences the assumed renewable power. At each time sample $\tM$ we record the current position $\posM^k$ and state of energy $E_\mathrm{b,M}^k$ of each aircraft $k$, the latter of which is assumed to only be available when the aircraft is charging on the ground, and the current state of energy $E_\mathrm{b,M}^h$ of the BESS at each airport $h$. Furthermore, we measure the currently available solar powers $P_\mathrm{rnw,M}^h$, from which we forecast future solar yield as explained in Section~\ref{subsec:airport}.

The controller is aimed at reducing the grid dependence of the airport network through the assignment and charge scheduling of airplanes for the given flight schedule, thus minimizing the objective function defined in Section~\ref{subsec:Objective}. A compelling future extension to this is in the context of airports as energy hubs, where they would supply power to their local communities while guaranteeing flight operations.
\subsection{Aviation Network Model}
We model the aviation network at the current time instance $\tM$ for a horizon of $N$ discrete time steps of $\Delta t$ as a time-varying, time-extended directed acyclic graph (DAG) $\mathcal{G}_t=(\mathcal{V}_t,\mathcal{A}_t)$. For brevity, we henceforth call the set of time steps in the horizon \mbox{$\mathcal{N} = \{\tM,\tM+\Delta t,\dots,\tM+N\cdot\Delta t \}$}. A vertex $i \in \mathcal{V}_t$ is the tuple $i=(a,t)$, with $i_\mathrm{s} = a \in \mathcal{H}$ and $i_\mathrm{t} = t \in \mathcal{N}$, i.e., there exists a node for every airport at every time step in the horizon. The set of edges $\mathcal{A}_t$ is constructed at every time step in the following way and as exemplified in Fig.~\ref{fig:DAG}:
Two vertices at the same airport and consecutive time steps are connected by ground edges, creating the set 
\par\nobreak\vspace{\vsp}\begin{small}\begin{equation*}
	\mathcal{A}_t^\mathrm{g} = \{(i,j): j_\mathrm{s} = i_\mathrm{s},\, j_\mathrm{t} = i_\mathrm{t}+1\} \; .
\end{equation*}\end{small}%
We extract all flights that start in the upcoming horizon from the given flight schedule and construct a set of flight edges that connect nodes of different airports. For every flight we allow a maximum possible flight delay of $d_\mathrm{max}$, an integer multiple of $\Delta t$, meaning that every flight adds $\frac{d_\mathrm{max}}{\Delta t}+1$ flight edges to the edge set, i.e., 
\par\nobreak\vspace{\vsp}\begin{small}\begin{align*}
	\mathcal{A}^f_t = &{\Big\{}(i,j):i_\mathrm{s}=o_f,\,j_\mathrm{s} =d_f,\, i_\mathrm{t} = t_f^\mathrm{sd} + \tau \; \forall \tau \in \{0,\dots,\frac{d_\mathrm{max}}{\Delta t}\},\\ & j_\mathrm{t} = i_\mathrm{t}+1 {\Big\}} \quad \forall f: t_f^\mathrm{sd} \in \mathcal{N}  \; .
\end{align*}\end{small}%
Note that the flight edges do not connect nodes that are $\frac{\hat{t}_f}{\Delta t}$ time steps apart but only span one time step, as this enables to track the energy per time step of each aircraft that traverses the graph. To ensure that aircraft need to respect the flight time for a flight assigned to them, there exists a set of virtual flight edges for every flight edge, that is
\par\nobreak\vspace{\vsp}\begin{small}\begin{align*}
		\nonumber \mathcal{C}_{(i,j)} = &\Big\{ (l,m) : \, l_\mathrm{s} = m_\mathrm{s} = j_\mathrm{s} , \, l_\mathrm{t} = j_\mathrm{t} + \tau - 1, \, m_\mathrm{t} = j_\mathrm{t} + \tau, \label{eq:Cij}\\
		&\forall \tau \in \{ 1,\ldots,t^f - 1 \}  \Big\} \quad \forall (i,j) \in \mathcal{A}^f_t, \, \forall f: t_f^\mathrm{sd} \in \mathcal{N}.
\end{align*}\end{small}%
With these virtual flight edges $\mathcal{C}_{(i,j)}\subset \mathcal{A}^\mathrm{g}_t$ we introduce path constraints for the aircraft in Section~\ref{subsec:aircraft}. Finally, to always ensure path feasibility, we introduce a terminal node $z$ and a set of virtual final edges 
\par\nobreak\vspace{\vsp}\begin{small}\begin{equation*}
		\mathcal{A}^\mathrm{v}_t = \{(i,z): i_\mathrm{t} = \tM+N\cdot\Delta t\}\quad  \forall i:i_\mathrm{s} \in \mathcal{H} \; , 
\end{equation*}\end{small}%
such that the complete edge set is defined as 
\par\nobreak\vspace{0pt}\begin{small}\begin{equation*}
		\mathcal{A}_t = \mathcal{A}^\mathrm{g}_t \cup \underset{f \in \mathcal{F}}{\bigcup}\mathcal{A}^f_t \cup \mathcal{A}^\mathrm{v}_t \, .
\end{equation*}\end{small}%
\begin{figure}[t!] 
	\centering
	\input{sections/dagFigure.tikz}
	\caption{Example of a DAG constructed at time $\tM$ for a horizon of $N$ time steps $\Delta t$ apart with 2 airports and 1 flight scheduled in the horizon. The virtual flight edges in the set of virtual flight edges corresponding to the first edge in $\mathcal{A}^f_t$ are highlighted in red. Aircraft 1 is on the ground at $\tM$, while aircraft 2 is en-route and thus its path origin is marked in the graph at its estimated time of arrival.\label{fig:DAG}}
\end{figure}
\subsection{Aircraft Routing \label{subsec:aircraft}}
In this section we establish the path and energy constraints for the aircraft as they are routed along the graph.  First, we introduce a binary variable $x_{(i,j)}^k$ for each aircraft and each edge, which is non-zero if that edge is part of the path. On ground edges the aircraft can recharge with a power $P_{(i,j)}^k$ as that means that they are at an airport. We concatenate all $x_{(i,j)}^k$ and $P_{(i,j)}^k$ into vectors $\mathbf{x}^k \in \{0,1\}^{|\mathcal{A}_t|\times 1}$ and $\mathbf{P}^k \in \mathbb{R}^{|\mathcal{A}_t^\mathrm{g}|\times 1}$, respectively, and introduce a vector $\mathbf{p}^k_t \in \{-1,0,1\}^{|\mathcal{V}_t|\times 1}$ that includes each aircraft's origin and destination node, i.e.,
\par\nobreak\vspace{-7pt}\begin{small}\begin{equation*}
		[\mathbf{p}^k_t]_i = \begin{cases}
			-1 \quad &i = \hat{p}_\mathrm{M}^k\\
			1 \quad &i = z\\
			0 \quad &\text{otherwise} 
		\end{cases} \quad \forall k \in \mathcal{P} \; ,
\end{equation*}\end{small}%
where $\hat{p}_\mathrm{M}^k$ is the estimated position of the plane. For planes that are on the ground at $\tM$ the position is known, i.e.,  $\hat{p}_\mathrm{M}^k=p_\mathrm{M}^k$, and their path origin is at one of the nodes with $i_\mathrm{t} = \tM$. Planes that are currently en-route, however, have their path origin instead placed at the node with $i_\mathrm{s}$ being their destination airport, and $i_\mathrm{t}$ their estimated time of arrival, as exemplified in Fig.~\ref{fig:DAG}.

The concatenated binary vector for all airplanes is denoted $\mathbf{x} = [\mathbf{x}^{1\top} \; \mathbf{x}^{2\top} \; \dots \; \mathbf{x}^{|\mathcal{P}|\top}]^\top$. Further, we denote $B_t$ the incidence matrix of $\mathcal{G}_t$ and represent the sets $\mathcal{C}_{(i,j)}$ with $C_t \in \{0,1\}^{|\mathcal{A}_t|\times |\mathcal{A}_t|}$ where
\par\nobreak\vspace{\vsp}\begin{small}\begin{equation*}
		[C_t]_{ab} = \begin{cases}
			1 \quad &\{b = (i,j),\, a=(l,m): a\in \mathcal{C}_b\}\\
			0 \quad &\text{otherwise} 
		\end{cases} \; ,
\end{equation*}\end{small}%
and an auxiliary matrix $M_t \in \{0,1\}^{|\mathcal{A}_t^\mathrm{g}|\times |\mathcal{A}_t|}$ that extracts ground edges from the edge set, i.e.,
\par\nobreak\vspace{\vsp}\begin{small}\begin{equation*}
		[M_t]_{ab} = \begin{cases}
			\bar{M} \quad & a=(i,j) \in \mathcal{A}_t,\,b = (i,j) \in \mathcal{A}_t^\mathrm{g}\\
			0 \quad &\text{otherwise} 
		\end{cases} \; ,
\end{equation*}\end{small}%
where $\bar{M}$ is a very large number in order to implement the big-M formulation which is used here to encode logic constraints~\cite{RichardsHow2005}. Furthermore, $A_t \in \{0,1\}^{|\{f:t_\mathrm{sd}^f \in \mathcal{N}\}|\times|\mathcal{A}_t|}$ exists if there are scheduled flights in the horizon and is defined as
\par\nobreak\vspace{\vsp}\begin{small}\begin{equation*}
		[A_t]_{fa} = \begin{cases}
			1 \quad & a=(i,j) \in \mathcal{A}_t^f\\
			0 \quad &\text{otherwise} 
		\end{cases} \; .
\end{equation*}\end{small}%
Then, the path constraints read	 
\par\nobreak\vspace{\vsp}\begin{small}\begin{align}
	B_t \, 	\mathbf{x}^k = \mathbf{p}^k_t \quad &\forall k \in \mathcal{P} \; , \label{eq:cont}\\
	\mathbf{x}^k \geq C_t \, \mathbf{x}^k  \quad &\forall k \in \mathcal{P} \; , \label{eq:xOnCij}\\
	A_t \, \mathbf{x} = \mathbf{1} \quad &\text{if } \exists f:t_\mathrm{sd}^f \in \mathcal{N} \; , \label{eq:flightsAssigned}\\
	\mathbf{P}^k \leq M_t\, \mathbf{x}^k \quad &\forall k \in \mathcal{P} \; , \label{eq:noChargeIfx=0}\\
	\mathbf{P}^k \leq M_t\,( \mathbf{1} - C_t \, \mathbf{x}^k) \quad &\forall k \in \mathcal{P} \; , \label{eq:noChargeOnCij}\\
	[\mathbf{P}^k]_i \in [P_\mathrm{c,min},P_\mathrm{c,max}] \quad \forall i \in \mathcal{A}_t^\mathrm{g} \; &\forall k \in \mathcal{P} \; . \label{eq:PcLimits}
\end{align}\end{small}%
Above, \eqref{eq:cont} enforces path continuity and \eqref{eq:xOnCij} ensures that all virtual flight edges are part of the path if their respective flight is assigned to $k$. By \eqref{eq:flightsAssigned} all flights in the horizon are ensured to be assigned. Through \eqref{eq:noChargeIfx=0} and \eqref{eq:noChargeOnCij} aircraft can only charge on ground edges that are part of their path, but not on virtual flight edges, as they are technically still flying during those time steps. The limits on the charging power are implemented in \eqref{eq:PcLimits}. 

A path is only physically possible if the aircraft does not run out of battery energy along it, which is why we carefully track the aircraft's state of energy at each time step through the state variable $E_\mathrm{b}^k[t]$. The energy evolves in between time steps through
\par\nobreak\vspace{\vsp}\begin{small}\begin{equation*}
	E_\mathrm{b}^k[t+1] = E_\mathrm{b}^k[t] - \underset{(i,j):i_\mathrm{t}=t\quad}{\sum P_{(i,j)}^k} \cdot \Delta t + \sum_{f\in\mathcal{F}}\underset{ (i,j)\in \mathcal{A}^f : i_\mathrm{t} = t\;\;}{\sum x_{(i,j)}^k\cdot \hat{E}_f }.	\label{eq:acBATdynamics}
\end{equation*}\end{small}%
We concatenate $E_\mathrm{b}^k[t]$ into $\mathbf{E}^k$ and introduce $G_t = \mathbf{1}_{|\mathcal{H}|}^\top \otimes \Lambda_N$ and $F_t \in \mathbb{R}^{|\mathcal{F}_t\times \mathcal{A}_t|}$, the latter of which only exists if there are scheduled flights in the horizon and is defined as
\par\nobreak\vspace{-7pt}\begin{small}\begin{equation*}
	[F_t]_{fa} = \begin{cases}
		\hat{E}_f \quad & a = (i,j) \in \mathcal{A}^f_t\\
		0 \quad &\text{otherwise}
	\end{cases} \; .
\end{equation*}\end{small}%
Therefore, the aircraft's battery constraints are introduced as
\par\nobreak\vspace{\vsp}\begin{small}\begin{align}
	\mathbf{E}^k = E_\mathrm{b,M}^k + G_t \, \mathbf{P}^k \cdot \Delta t - F_t\, \mathbf{x}^k \quad &\forall k \in \mathcal{P} \; ,\label{eq:Ebk(t)} \\
	\mathbf{E}^k \geq [ \mathbf{1}_{|\mathcal{H}|}^\top \otimes I_N] \, M_t\, C_t\,\mathbf{x}^k\cdot E_\mathrm{res} \quad &\forall k \in \mathcal{P} \; , \label{eq:Ereserve} \\
 	[\mathbf{E}^k]_i \in [E_\mathrm{b,min}^k,E_\mathrm{b,max}^k] \quad \forall i \in \mathcal{N} \; &\forall k \in \mathcal{P} \; , \label{eq:EbkLimits}\\
 	[\mathbf{E}^k]_N \in \mathcal{X}_\mathbf{f}^{E_\mathrm{b}} \quad &\forall k \in \mathcal{P} \; , \label{eq:acBatFinalSet}
\end{align}\end{small}%
where \eqref{eq:Ebk(t)} implements the battery dynamics, \eqref{eq:Ereserve} ensures that the aircraft has an energy reserve $E_\mathrm{res}$ after the flight, and \eqref{eq:EbkLimits} constrains the battery energy to stay within its bounds. The state of energy in the battery at the final time step of the horizon is constrained to be in a terminal set through \eqref{eq:acBatFinalSet}.
\subsection{Energy Model of the Airport \label{subsec:airport}}
The airport's energy system is modeled with an on-site BESS and a connection to both renewable energy sources and the grid as in Fig.~\ref{fig:airport}. First, we introduce \mbox{$\mathbf{P}_\mathrm{c} = [\mathbf{P}^{1\top} \; \mathbf{P}^{2\top} \; \dots \; \mathbf{P}^{|\mathcal{P}|\top}]^\top$} which collects the vectors for the charging power of each aircraft, and denote the battery energy and power of the BESS as $\mathbf{E}_\mathrm{b}^h$ and $\mathbf{P}_\mathrm{b}^h$, respectively. At each airport we forecast the solar power per time step over the upcoming horizon from the previous time steps, which we concatenate in $\hat{\mathbf{P}}_\mathrm{rnw}^h$. For a first implementation we adopt a Holt-Winters forecasting model~\cite{Holt2004}, which can be changed for higher fidelity forecasting methods in the future.
The airport-specific power and energy constraints are defined as
\par\nobreak\vspace{\vsp}\begin{small}\begin{align}
	\mathbf{E}_\mathrm{b}^h = E_\mathrm{b,M}^h - \Lambda_N \, \mathbf{P}_\mathrm{b}^h\cdot \Delta t \quad &\forall h \in \mathcal{H}\; , \label{eq:BESSdynamics}\\
	\mathbf{P}_\mathrm{gr}^h \geq \mathbf{P}_\mathrm{a}^h - \hat{\mathbf{P}}_\mathrm{rnw}^h - \mathbf{P}_\mathrm{b}^h \quad &\forall h \in \mathcal{H}\; , \label{eq:gridPower}\\
	\mathbf{P}_\mathrm{a}^h = [\mathbf{1}_{|\mathcal{P}|}^\top \otimes I_{|\mathcal{H}|\cdot N}]\, \mathbf{P}_\mathrm{c} \quad &\forall h \in \mathcal{H} \; ,\label{eq:apronPower}\\
	\mathbf{E}_\mathrm{b}^h \in [E_\mathrm{b,min}^h,E_\mathrm{b,max}^h] \quad &\forall h \in \mathcal{H} \;. \label{eq:EbhLimits}
\end{align}\end{small}%
The battery dynamics are enforced through \eqref{eq:BESSdynamics} and the grid power is obtained through a power balance in \eqref{eq:gridPower}, where the apron power is defined in \eqref{eq:apronPower}. The BESS energy is confined to limits in \eqref{eq:EbhLimits}.
\subsection{MPC Formulation\label{subsec:Objective}}
Since the goal of the controller is to minimize grid dependence while guaranteeing an acceptable level of service in the flight network, we formulate a multi-objective cost function that penalizes the use of grid power and flight delays. In addition to that, we add a terminal cost for the final state of charge in the BESS such that it is kept as high as possible. We introduce $\mathbf{q}$ that contains the penalty terms for the grid power, and $\mathbf{r}_t$, with
\par\nobreak\vspace{\vsp}\begin{small}\begin{equation*}
		[\mathbf{r}_t]_a = \begin{cases}
			\varphi\cdot(i_\mathrm{t}-t_\mathrm{sd}^f)^2 \quad & a=(i,j):(i,j)\in\mathcal{A}^f_t\\
			0 \quad &\text{otherwise} 
		\end{cases} \; ,
\end{equation*}\end{small}%
where $\varphi$ is a weighting factor. Finally, we define a weight $\vartheta$ on the terminal cost and define the receding-horizon optimization problem as \smallskip

\begin{prob}{Aircraft Routing and Charge Scheduling (ARCS)}\label{prob:MPC} The optimal aircraft assignment for the horizon is found through the solution of
	\begin{align*}
		\min_{\mathbf{x},\mathbf{P},\{\mathbf{P}_\mathrm{b}^h\}_{h\in\mathcal{H}}} \quad & \sum_{h\in\mathcal{H}} \mathbf{q}^\top\, \mathbf{P}_\mathrm{gr}^h + \mathbf{r}_t^\top\,\mathbf{x} - \vartheta \cdot \sum_{h\in\mathcal{H}} E_{\mathrm{b},N}^h\\
		\begin{aligned}
			&\mathrm{s.t.}\\&\\&
		\end{aligned} \quad&\begin{aligned}
			 &\eqref{eq:cont}-\eqref{eq:PcLimits}\;\;&\textit{Path Constraints},\\
					& \eqref{eq:Ebk(t)}-\eqref{eq:acBatFinalSet}\;\;&\textit{Aircraft Constraints},\\
					& \eqref{eq:BESSdynamics}-\eqref{eq:EbhLimits}\;\;&\textit{Airport Constraints}.
				\end{aligned}
	\end{align*}
\end{prob}%
We implement Problem~1 in the receding horizon control framework shown in Fig.~\ref{fig:mpc}.
\begin{figure}[t]
	\centering
	\input{sections/arcs}
	\caption{Scheme for implementation of the Aircraft Routing and Charge Scheduling (ARCS). Blocks with rounded edges are exogenous inputs and variables refer to the respective quantity at the measured time unless otherwise specified.}\label{fig:mpc}
\end{figure}

%% file: sections/dagFigure.tikz
	\begin{tikzpicture}[
	groundEdge/.style = {draw=black, line width= 1pt, -{Triangle[fill=black]}},
	flightEdge/.style = {draw=black!50, line width= 1pt, -{Triangle[fill=black!50]}},
	virtualEdge/.style = {dashed,draw=black!50, line width= 1pt, -{Triangle[fill=black!50]}},
	ij/.style = {draw=red, line width= 1pt, -{Triangle[fill=red]}},   
	Cij/.style = {draw=red!40, line width= 4pt, -},
	pos1/.style = {draw=ForestGreen, line width= 1pt, -{Triangle[fill=ForestGreen]}}, 
	pos2/.style = {draw=YellowOrange, line width= 1pt, -{Triangle[fill=YellowOrange]}}, 
	]
	\pgfmathtruncatemacro{\h}{1.5};
	\node (gE) at (.2,-.9) [anchor=west] {$\in \mathcal{A}^\mathrm{g}_t$};
	\node (fE) at (1.7,-.9) [anchor=west] {$\in \mathcal{A}^f_t$};
	\node (vE) at (3.2,-.9) [anchor=west] {$\in \mathcal{A}^\mathrm{v}_t$};
	\draw[groundEdge] ($(gE)+(-1,0)$) -- (gE);
	\draw[flightEdge] ($(fE)+(-1,0)$) -- (fE);
	\draw[virtualEdge] ($(vE)+(-1,0)$) -- (vE);
	\node (origin) at (4.5,-.9) [anchor=west] {origin};
	\node[fill=gray,circle,minimum width=3pt,inner sep=2pt] at ($(origin.west)+(-.1,-.1)$) (orig) {};
	\draw[draw=gray, line width= 1pt, -{Triangle[fill=gray]}] ($(orig)+(0,0.3)$) -- ($(orig)+(0,0)$);
	\node (dest) at ($(origin.east)+(.2,0.02)$) [anchor=west] {destination};
	\node[fill=gray,circle,minimum width=3pt,inner sep=2pt] at ($(dest.west)+(-.1,-.1)$) (desti) {};
	\draw[draw=gray, line width= 1pt, -{Triangle[fill=gray]}] ($(desti)$) -- ($(desti)+(0,0.3)$);
	\draw (-0.4,-.65) rectangle ($(dest.east)+(.1,-.3)$);
	\draw[dashed] (0,\h+0.2) -- (0,-.2);
	\draw[dashed] (6,\h+0.2) -- (6,-.2);
	\node (t) at (0,-0.4) {$\tM$};
	\node (t1) at (1.2,-0.4) {$\tM+\Delta t$};
	\node (tN) at (6.2,-0.4) {$\tM+N\cdot \Delta t$};
	\node (zL) at (7.2,\h/2-.1) {$z$};
	\node (std) at (2,\h+0.4) {$t_f^\mathrm{sd}$};
	\draw (2,\h+0.8) edge[|-|] node[fill=white,inner sep=2pt] {$d_\mathrm{max}$} (3,\h+0.8);
	\draw (2,\h+1) edge[red,|-|] node[fill=white,inner sep=2pt] {\color{red}$\hat{t}_f$} (5,\h+1);
	\begin{scope}[every node/.style={circle,minimum width=5pt,thick,draw,inner sep=2pt,fill=white}]
		\foreach \i in {0,...,6}
		{
			\pgfmathtruncatemacro{\label}{\i*2+1};
			\node (\label) at (\i,\h) {};
			\pgfmathtruncatemacro{\label}{\i*2+2};
			\node (\label) at (\i,0) {};
		}
		\node (z) at (7,\h/2) {};
	\end{scope}
	\draw[Cij] (8) edge (10);
	\draw[Cij] (10) edge (12);
	\node (setCij) at (3.8,-.4) {\color{red!60}$\mathcal{C}_{(i,j)}$};
	\foreach \i in {0,...,5}
	{
		\pgfmathtruncatemacro{\labelIn}{\i*2+1};
		\pgfmathtruncatemacro{\labelOut}{(\i+1)*2+1};
		\draw[groundEdge] (\labelIn) edge (\labelOut);
		\pgfmathtruncatemacro{\labelIn}{\i*2+2};
		\pgfmathtruncatemacro{\labelOut}{(\i+1)*2+2};
		\draw[groundEdge] (\labelIn) edge (\labelOut);
	}
	\draw[flightEdge] (7) edge (10);
	\draw[flightEdge] (5) edge node[ellipse,fill=white,inner sep=0pt] {\footnotesize\color{red}$(i,j)$} (8);
	\draw[virtualEdge] (13) -- (z);
	\draw[virtualEdge] (14) -- (z);
	\node[fill=ForestGreen,circle,minimum width=3pt,inner sep=2pt] (p1) at (0,\h+.05) {};
	\node[fill=YellowOrange,circle,minimum width=3pt,inner sep=2pt] (p2) at (2,0.05) {};
	\node[anchor=west] (p1L) at (0,\h+.4) {\footnotesize\color{ForestGreen}$\posM^1$};
	\node[anchor=east] (p2L) at (2,.4) {\footnotesize\color{YellowOrange}$\hat{p}_\mathrm{M}^2$};
	\draw[pos1] (0,\h+.4) -- (0,\h+.05);
	\draw[pos2] (2,.4) -- (2,0.05);
	\node[fill=ForestGreen,circle,minimum width=3pt,inner sep=2pt] (e1) at (6.95,\h/2+.05) {};
	\node[fill=YellowOrange,circle,minimum width=3pt,inner sep=2pt] (e2) at (7.05,\h/2+.05) {};
	\draw[pos1] (6.95,\h/2) -- (6.95,\h/2+.4);
	\draw[pos2] (7.05,\h/2) -- (7.05,\h/2+.4);
\end{tikzpicture}

%% file: sections/arcs.tex
\begin{tikzpicture}[
	node distance = 4mm and 15mm,
	block/.style = {draw, minimum width=6mm, font={\footnotesize},fill=white,inner sep=3pt},
	arr/.style = {draw=black!60, line width= 1pt, -{Triangle[fill=black!60]}},
	arrlabelHorz/.style = {anchor=south,font={\tiny}},
	arrlabelVert/.style = {anchor=west,font={\tiny}},
	]
	\node[block,minimum height=15mm,thick] (mpc) {$\;\;$\textbf{ARCS}$\;\;$};
	\node[block,below right =2mm and 15mm of mpc.east,minimum height=8mm] (fleet) {$\quad\qquad$Fleet};
	\node[block,below right=1mm and 20mm of mpc.north east] (h1) {\phantom{Airports}};
	\node[block] at ($(h1)+(-1mm,1mm)$) {\phantom{Airports}};
	\node[block] at ($(h1)+(-2mm,2mm)$) (airports) {Airports};
	\node[block,above left = 7mm and 5mm of mpc.north west,anchor=south] (forecast) {\begin{tabular}{c}
			solar\\ forecast
	\end{tabular}};
	\node[block,rounded corners=1mm,above=7mm of airports.north] (solar) {real solar power};
	\node[below=9mm of forecast.south] (auxSolar) {};
	\node[block,left=of mpc.west] (graph) {\begin{tabular}{c}
			build\\ graph
	\end{tabular}};
	\node[block,rounded corners=1mm,left=2mm of forecast.west] (schedule) {schedule};
	\draw[arr] (mpc.east |- fleet.west) to node[arrlabelHorz,xshift=-1pt] {\begin{tabular}{l}
			flight assignments,\\
			$\{P^k\}_{k\in\mathcal{P}}$
	\end{tabular}} (fleet.west);
	\draw[arr] (mpc.east |- airports.west) to node[arrlabelHorz] {$\{P_\mathrm{b}^h,P_\mathrm{gr}^h\}_{h\in\mathcal{H}}$} (airports.west);
	\draw[arr] (fleet.north -| airports.south) to node[arrlabelVert,yshift=-1mm] {$\{P_\mathrm{a}^h\}_{h\in\mathcal{H}}$} (airports.south);
	\draw[arr] (solar.south) to node[arrlabelVert] {$\{P_\mathrm{rnw}^h\}_{h\in\mathcal{H}}$} (airports.north);
	\draw[arr] (solar.west) to node[arrlabelHorz] {$\{P_\mathrm{rnw}^h\}_{h\in\mathcal{H}}$} (solar.west -| forecast.east);
	\draw[arr] (forecast.south) to node[anchor=west,yshift=2mm,xshift=-1mm,font={\tiny}] {$\{\hat{P}_\mathrm{rnw}^h[t]\}_{h\in\mathcal{H},\atop t\in\mathcal{N}}$} (auxSolar.center) to (auxSolar.center -| mpc.north west);
	\draw[arr] (schedule.south -|graph.north) to node[arrlabelVert,xshift=-1mm] {$\{f:t_\mathrm{sd}^f\in\mathcal{N}\}$} (graph.north);
	\draw[arr] (graph.east) to node[anchor=south,xshift=-2mm,font={\tiny}] {$\mathcal{G}_t,\mathrm{p}_t$} (mpc.west);
	\node[below=4mm of mpc.south] (auxFleet1) {};
	\node[above left=1mm and 5mm of mpc.south west] (auxFleet2) {};
	\draw[arr] (fleet.south) to (fleet.south |- auxFleet1.center) to (auxFleet1.center -| forecast.south) to node[arrlabelVert,xshift=-1mm,yshift=-1mm] {$\{E_\mathrm{b}^k,\hat{p}_\mathrm{M}^k\}_{k\in\mathrm{P}}$} (forecast.south |- auxFleet2.center) to (auxFleet2.center -| mpc.west);
	\node[right=5mm of fleet.east] (auxAirport1) {};
	\node[below=5mm of mpc.south] (auxAirport2) {};
	\node[above left=3mm and 6mm of mpc.south west] (auxAirport3) {};
	\draw[arr] (airports.east) to (airports.east -| auxAirport1.center) to (auxAirport1.center |- auxAirport2.center) to (auxAirport2.center -| auxAirport3.center) to node[anchor=east,font={\tiny}] {$\{E_\mathrm{b}^h\}_{h\in\mathcal{H}}$} (auxAirport3.center) to (auxAirport3.center -| mpc.west); 
	\begin{scope}[shift={($(fleet.west)+(5mm,-5mm)$)},scale=0.25]
		\draw[fill=white] (0,2) -- (-1.7,1.8) -- (-1.7,1.6) -- (0,1.5) -- (1.7,1.6) -- (1.7,1.8) -- cycle;
		\draw[fill=white,x radius = .6,y radius=.1] (0,.2) circle;
		\draw[rounded corners=3,fill=white] 
		(.05,.05) .. controls (0,0) .. (-.05,.05)
		(-.05,.05) .. controls  (-0.7,2) and (-.1,3) .. (0,3.2)
		.. controls (.1,3) and (.7,2) .. (.05,.05);
	\end{scope}
	\begin{scope}[shift={($(fleet.west)+(3mm,-3.5mm)$)},scale=0.25]
		\draw[fill=white] (0,2) -- (-1.7,1.8) -- (-1.7,1.6) -- (0,1.5) -- (1.7,1.6) -- (1.7,1.8) -- cycle;
		\draw[fill=white,x radius = .6,y radius=.1] (0,.2) circle;
		\draw[rounded corners=3,fill=white] 
		(.05,.05) .. controls (0,0) .. (-.05,.05)
		(-.05,.05) .. controls  (-0.7,2) and (-.1,3) .. (0,3.2)
		.. controls (.1,3) and (.7,2) .. (.05,.05);
	\end{scope}
	\begin{scope}[shift={($(fleet.west)+(1mm,-2mm)$)},scale=0.25]
		\draw[fill=white] (0,2) -- (-1.7,1.8) -- (-1.7,1.6) -- (0,1.5) -- (1.7,1.6) -- (1.7,1.8) -- cycle;
		\draw[fill=white,x radius = .6,y radius=.1] (0,.2) circle;
		\draw[rounded corners=3,fill=white] 
		(.05,.05) .. controls (0,0) .. (-.05,.05)
		(-.05,.05) .. controls  (-0.7,2) and (-.1,3) .. (0,3.2)
		.. controls (.1,3) and (.7,2) .. (.05,.05);
	\end{scope}
\end{tikzpicture}

%% file: sections/results.tex
\section{Numerical Simulations}
\begin{figure}[b!]
	\centering
	\includegraphics[width=\columnwidth]{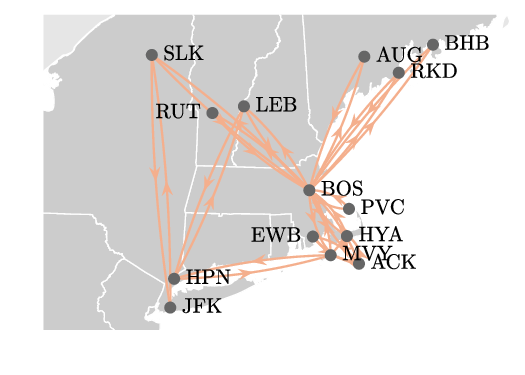}
	\caption{Flight network operated by Cape Air in New England.\label{fig:network}}
\end{figure}

We simulate our control framework on Cape Air's commuter flight network in New England, where the airline operates around 200 daily flights between 14 airports as shown in Fig.~\ref{fig:network}. We use flights from a week in August 2024~\cite{Menger2025} and obtain solar irradiation data in 10\,min increments for every airport through the National Solar Radiation Data Base (NSRDB)~\cite{SenguptaXieEtAl2018}.

We assume that every airport in the network has 1000\,m$^2$ of solar cells and a 500\,kWh BESS. We parse Problem~1 with YALMIP~\cite{Loefberg2004} and solve it with Gurobi~\cite{GurobiOptimization2021} with a horizon of 2\,h and a discrete time step of 5\,min, resulting in an average computation time of few seconds for $|\mathcal{A}_t|\cdot |\mathcal{P}| \approx 15,000$ binary variables per horizon.
\subsection{Flight Simulations}
To simulate the scheduled flights between airports we adopt an agent-based approach with simplified quasi-static 2-D flight dynamics for every aircraft. Due to the absence of commercial electric aircraft at the time of writing, we envisage an electric version of the P2012 Traveler, inspired by~\cite{JustinEtAl2017} as the aircraft for our case study.

When a flight is assigned to an aircraft, a distance-based trajectory is computed that starts with a steady climb at a fixed angle to 1000\,m, followed by a cruise phase at constant velocity and a descent once a certain way point before the destination airport is reached. The aircraft follows this trajectory as closely as possible, and every increment of time $\delta t$, the current thrust requirements are obtained through
\begin{small}
\begin{equation*}
	F_\mathrm{t}[k] = \frac{1}{2}\rho[k] S \left(c_\mathrm{D,0}+\frac{c_\mathrm{L}^2[k]}{\pi AR e} \right) v^2[k] + m \left(\sin(\theta[k]) g + a[k]\right)\; ,
\end{equation*}\end{small}%
where $\rho[k]$ is the density at the current altitude, $S$ is the wing area, $c_\mathrm{D,0}$ the parasitic drag, $AR$ and $e$ the plane's aspect ratio and Oswald efficiency, $\theta[k]$ the current climb angle, $v[k]$ and $a[k]$ are the velocity and acceleration in direction of travel, $m$ the mass and $g$ the gravitational constant~\cite[Ch's.~17-22]{Gudmundsson2014}. On the ground, an additional drag term is added, namely $(m g - \nicefrac{1}{2}\rho[k]S c_\mathrm{L}[k]v^2[k])\mu$, where the friction coefficient $\mu$ includes braking during landing. The lift coefficient is 
\begin{equation*}
	c_\mathrm{L}[k] = \frac{2 m g \cos(\theta[k])}{S \rho[k] v^2[k]} \; .
\end{equation*}
We assume that during flight the airplane stays within its operational envelope, but restrict the deceleration during descent and landing such that the required lift coefficient does not exceed the maximum to prevent stalling. The propeller efficiency is interpolated from a look-up table
\begin{equation*}
	\eta_\mathrm{prop}[k] = \mathcal{M}\left(P_\mathrm{t}[k],\rho[k],v[k]\right) \; ,
\end{equation*}
where $P_\mathrm{t}[k] = F_\mathrm{t}[k] v[k]$ is the thrust power (cf.~\cite[Ch.~14]{Gudmundsson2014}). For the battery dynamics we assume a constant (dis-)charging efficiency, such that the state of energy can be obtained as
\begin{equation*}
	E_\mathrm{b}[k+1] =  \begin{cases}
		E_\mathrm{b}[k] - \frac{1}{\eta_\mathrm{b}\eta_\mathrm{prop}[k]}  P_\mathrm{t}[k]\delta t \;& \text{during flight,}\\
		E_\mathrm{b}[k] +\eta_\mathrm{b} P_\mathrm{c}[k]\delta t \;\; &\text{when charging.}
	\end{cases}
\end{equation*}
We simulate the dynamics with $\delta t = 1\,\text{s}$ and run the MPC whenever the clock time is a multiple of $\Delta t$. 

\subsection{Results}
We show selected results for the case study at hand. First, we generate a baseline schedule by solving a fleet assignment problem similar to~\cite{Mitici2022,VehlhaberSalazar2024} which we use to assign aircraft to their initial locations. We obtain grid power requirements for said baseline, where we do not allow for aircraft reassignment, but only energy management, which we optimize with our framework by enforcing its binary variables to follow the baseline schedule, although this baseline control scheme could be achieved with even simpler energy management schemes.

\begin{figure}[t!]
	\centering
	\includegraphics[width=\columnwidth]{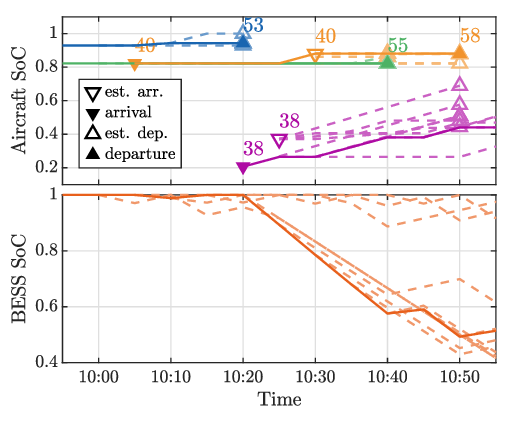}
	\caption{Comparison of projected (muted,dashed)  and actual (solid) state of charge (SoC) trajectories for a selected time window at Martha's Vineyard Airport (MVY). Flights 40/55 are from/to HPN, 40/(53,58) from/to BOS. \label{fig:mpcExample}}
\end{figure}

The proposed ARCS framework is applied as outlined in Fig.~\ref{fig:mpc}. Fig.~\ref{fig:mpcExample} shows an example for a successful re-assignment: Initially, the plane flying flight 38 was then projected to be assigned to flight 58 after a period of recharging. However, the aircraft arrives earlier and at a lower SoC than expected, prompting a re-assignment of flight 58 to the yellow aircraft for which it is charged a little more and swaps flights with another aircraft that was already on the tarmac. Potentially, since flight 58 is a short flight to Boston (BOS), a higher SoC saves grid power there at a later time.

The BESS at the airport is used for charging the purple aircraft, which it was for the longest time not projected to do. This is likely caused by a change in solar irradiation conditions, and its rather unstable unstable prediction, which will be addressed in future work.

At airports with many aircraft present within a horizon, as is often the case at BOS, we find that frequent re-scheduling can take place, which may compromise passenger level of service. We plan to address this issue by exploring objective functions that penalize schedule deviations.

\begin{figure}[t!]
	\centering
	\includegraphics[width=\columnwidth]{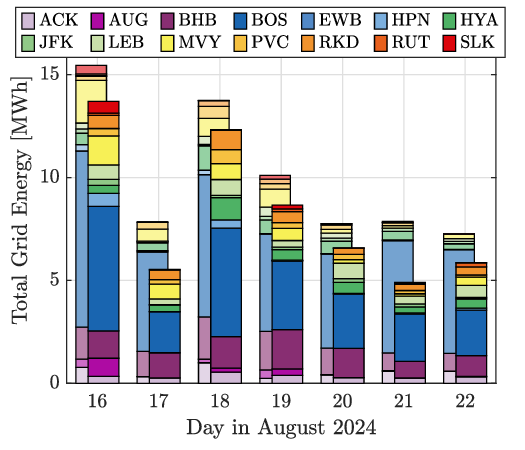}
	\caption{Comparison of grid energy requirements for electric aircraft operations at airports (with IATA identifiers) in the Cape Air flight network without (muted, background) and with (foreground) allowing for aircraft re-assignment.\label{fig:resBaseline}}
\end{figure}

	Fig.~\ref{fig:resBaseline} shows results for grid energy requirements for every day in the considered week compared to the baseline. Depending on schedule and weather conditions, we can save between 10~and~37~\% of total grid energy in the whole network, thus making better use of the renewable energy sources installed at the airports, notably without causing flight delays. We find a large proportion of these savings at BOS, which is the airline's main hub, with more than 80~\% of their flights either starting or landing there. At other airports, grid energy requirements rise slightly with the proposed scheme in some cases, which may be owed to the objective function that merely penalizes the use of total grid power without taking local circumstances into account. These findings motivate investigation into objective functions using (forecast) energy costs~\cite{FernandezZapicoHofmanEtAl2025} and sizing considerations for the energy infrastructure at specific airports.

%% file: sections/conclusion.tex
\FloatBarrier
\section{Conclusion}
\vspace{-5pt}
In this paper we introduced a model predictive control (MPC) scheme for dynamic flight re-assignment and energy management of electric regional aviation networks in which the airports are largely grid-independent.
Framed as a time-varying time-extended digraph, we combined a network-flow formulation with a linear energy system model and instantiated an MPC framework in order to compute aircraft assignment and energy trajectories in real-time.
Results on a network of commuter flights showed that changes to schedule and power allocation in real-time can successfully minimize disruptions in the schedule and ensure efficient operation.

Future work will add demand response for the local power grid to the framework, paving the way towards regional airports to serve as energy hubs. Furthermore, a stochastic approach may improve robustness and concepts from hybrid systems theory could help derive requirements for the terminal set. Another compelling research direction is to add concepts from ride-pooling~\cite{PaparellaPedrosoEtAl2024b} and passenger recapturing.